\documentclass{naturecomm}

\title{Spontaneous condensation of exciton polaritons \\ in the single-shot regime}

\author{E. Estrecho$^{1}$, T. Gao$^{1}$, N. Bobrovska$^{2}$, M. D. Fraser$^{3}$, M. Steger$^{4}$, L. Pfeiffer$^{5}$, K. West$^{5}$, \\ T. C. H. Liew$^{6}$, M. Matuszewski$^{2}$, D. W. Snoke$^{4}$, A. G. Truscott$^{7}$ \& E. A. Ostrovskaya$^{1}$}

\begin{document}

\maketitle

\begin{affiliations}
 \item Nonlinear Physics Centre, Research School of Physics and Engineering, The Australian National University, Canberra, ACT 2601, Australia
 \item Institute of Physics, Polish Academy of Sciences, A. Lotinik\'ow 32$/$46, 02-668 Warsaw, Poland
 \item Quantum Functional System Research Group, RIKEN Center for Emergent Matter Science, 2-1 Hirosawa, Wako-shi, Saitama 351-0198, Japan
 \item Department of Physics and Astronomy, University of Pittsburgh, PA 15260, USA
 \item Department of Electrical Engineering, Princeton University, Princeton, New Jersey 08544, USA
 \item Division of Physics and Applied Physics, Nanyang Technological University, Singapore
  \item Laser Physics Centre, Research School of Physics and Engineering, The Australian National University, Canberra, ACT 2601, Australia
\end{affiliations}

\begin{abstract}

Bose-Einstein condensate of exciton polaritons in a semiconductor microcavity is a macroscopically populated coherent quantum state subject to concurrent pumping and decay. Debates about the fundamental nature of the condensed phase in this open quantum system still persist. Here, we gain a new insight into the spontaneous condensation process by imaging {\em long-lifetime} exciton polaritons in a high-quality {\em inorganic} microcavity in the {\em single-shot} optical excitation regime, without averaging over multiple condensate realisations. In this highly non-stationary regime, a condensate is strongly influenced by the `hot' incoherent reservoir, and reservoir depletion is critical for the transition to the ground energy and momentum state. Condensates formed by more photonic exciton polaritons exhibit dramatic reservoir-induced density filamentation and shot-to-shot fluctuations. In contrast, condensates of more excitonic quasiparticles display smooth density and are second-order coherent. Our observations show that the single-shot measurements offer a unique opportunity to study formation of macroscopic phase coherence during a quantum phase transition in a solid state system.


\end{abstract}

Exciton polaritons (polaritons herein) in semiconductor microcavities with embedded quantum wells (QWs) \cite{Microcavity_book} are composite bosonic quasiparticles that result from strong coupling between cavity photons and excitons, and exhibit a transition to quantum degeneracy akin to Bose-Einstein condensation (BEC) \cite{Deng_02,BEC06,BEC07,Deng_10,CiutiREV13,YamamotoREV14}. All signatures of BEC, such as macroscopic occupation of a ground state, long-range coherence, quantised circulation \cite{vortices,vortex_pair}, and superfluidity \cite{superfluidity} have been observed in this system during the past decade. However, due to the inherent open-dissipative nature of the system stemming from the short lifetime of polaritons (from $\sim 10^1$ to $\sim 10^2$ ps) and the need to replenish them via optical or electrical pumping, the nature of the transition to the macroscopically occupied quantum state in polariton systems remains the subject of continuing debate. In particular, it has been conjectured that exciton-polaritons exhibit a highly non-equilibrium Berezinskii--Kosterlitz--Thouless (BKT) rather than BEC phase \cite{Roumpos12,BKT_PRX,BKT_arXiv}, while other studies support the assertion of exciton-polariton condensation at thermodynamic equilibrium \cite{BEC_equilibrium,Snoke}. The difficulty in resolving the nature of condensation lies in the short time scales of the polariton dynamics. Continuous-wave ({\em cw}) regime experiments deal with a {\em steady state} reached by this open-dissipative system, while pulsed experiments on polaritons are usually done by averaging over {\em millions} of realisations of the experiment. In both regimes, time-integrated imaging of the condensate by means of cavity photoluminescence  spectroscopy washes out any dynamical and stochastic processes. For example, proliferation of spontaneously created dynamic phase defects (vortices) and formation of the topological order in the BKT phase transition \cite{BKT_arXiv} cannot be unambiguously confirmed since only stationary vortices pinned by impurities or lattice disorder potential survive the averaging process. To understand the process of polariton condensation, and the evolution thereafter, one requires {\em single-shot imaging of condensation dynamics}. 

Another difficulty in interpreting the experimental results lies in the strong influence of a reservoir of incoherent, high-energy excitonic quasiparticles on the condensation dynamics due to the spatial overlap between the reservoir and condensing polaritons. This overlap is particularly significant when the condensate is created by a Gaussian-shaped optical pump spot, which can trap both the long-lived excitonic reservoir and the short-lived polaritons in the gain region \cite{multimode1,multimode2,gain_guiding_exp}. The interaction between the condensate and the reservoir particles is strong and repulsive, which enables creation of effective potentials by exploiting a local reservoir-induced energy barrier (blue shift) \cite{Bloch10} and a spatially structured optical pump. This technique has enabled observation of polariton condensates in a variety of optically-induced trapping geometries\cite{Bloch10,Tosi12,Tosi_vortices,Cristofolini13,Askitopoulos13,Dall14,Gao15,Liew_ring,Bayer15}, as well as creation of condensates effectively decoupled from the pump (reservoir) region \cite{Bloch10,Bloch11,SnokePRX13,Nelson15,Sanvitto_arXiv}. Despite the significant advances in creating and manipulating polariton condensates with the help of a steady-state reservoir induced by a {\em cw} pump, the influence of the reservoir on the condensate formation process is poorly understood. Recently, however, single-shot imaging performed on organic microcavities \cite{Bobrovska2016}, provided evidence in support of earlier theoretical suggestions that the reservoir is responsible for dynamical instability and subsequent spatial fragmentation of the polariton condensate in a wide range of excitation regimes \cite{Wouters2007,Smirnov14,Bobrovska14,Liew15}. Whether or not this behaviour is unique to organic materials, which are strongly influenced by material disorder, can only be determined by single-shot imaging in inorganic microcavities. Although single-shot experiments were previously performed in GaN and CdTe heterostructures \cite{Baumberg08,deveaud_g2}, single-shot real-space imaging of the condensate {\em was thought impossible in inorganic microcavities} due to insufficient brightness of the cavity photoluminescence.

In this work, we perform, for the first time to our knowledge, {\em single-shot real-space imaging} of exciton polaritons created by a short laser pulse in a high-quality inorganic microcavity supporting long-lifetime ($\sim 200$ ps) polaritons \cite{Snoke,ballistic,SnokePRX13}. The long lifetime allows for sufficient photoluminescence to be collected during the single-pulse experiment. Using this technique, we show that spatial fragmentation (filamentation) of the condensate density is an inherent property of a non-equilibrium, spontaneous Bose-Einstein condensation resulting from initial random population of high-energy and momenta states, and will persist even after relaxation to the lowest energy and momentum occurs. We unambiguously link this behaviour to the highly non-stationary nature of the condensate produced in a single-shot experiment, as well as to trapping of condensing polaritons in an effective random potential induced by spatially inhomogeneous depletion of the reservoir. Furthermore, we observe a transition from a  condensate with strong filamentation and large shot-to shot density fluctuations to a more homogeneous state with reduced density fluctuations. The two regimes of polariton condensation are accessible in our experiments due to a wide range of detuning between the cavity photon and QW excitons \cite{ballistic}. A negative detuning results in more photonic, light polaritons while positive detuning supports more excitonic, heavy quasiparticles exhibiting stronger interactions \cite{Microcavity_book,Deng_10}. We show that, while in the quasi-stationary  {\em cw} regime transition to ground-state condensation is driven by non-radiative (e.g., phonon-assisted \cite{SavenkoPRL}) energy relaxation processes that are more efficient for exciton-like polaritons \cite{two_regimes}, in the highly non-stationary single-shot regime this transition is driven primarily by reservoir depletion. 

\section*{Results}
\subsection{Transition to condensation.}
Spontaneous Bose-Einstein condensation of exciton polaritons is typically achieved with an optical pump which is tuned far above the exciton resonance in the microcavity \cite{BEC06}. The phonon-assisted and exciton-mediated relaxation of the injected free carriers \cite{Porras02} then efficiently populates the available energy states of the lower polariton (LP) dispersion branch $E(\bf{k})$, where $k$ is the momentum in the plane of the quantum well. The reduced efficiency of the relaxation processes leads to accumulation of the polaritons in the bottleneck region at a high energy close to that of the exciton \cite{CiutiREV13,Tassone97}. When stimulated scattering from this incoherent, high-energy excitonic reservoir into the ${\bf k}=0$  takes place, transition to condensation in the ground state of the LP dispersion $E_{\rm min}({\bf k}=0)$ is achieved \cite{BEC06,Tartakovskii00}. 

\begin{figure}
\centering
\includegraphics[width=13cm]{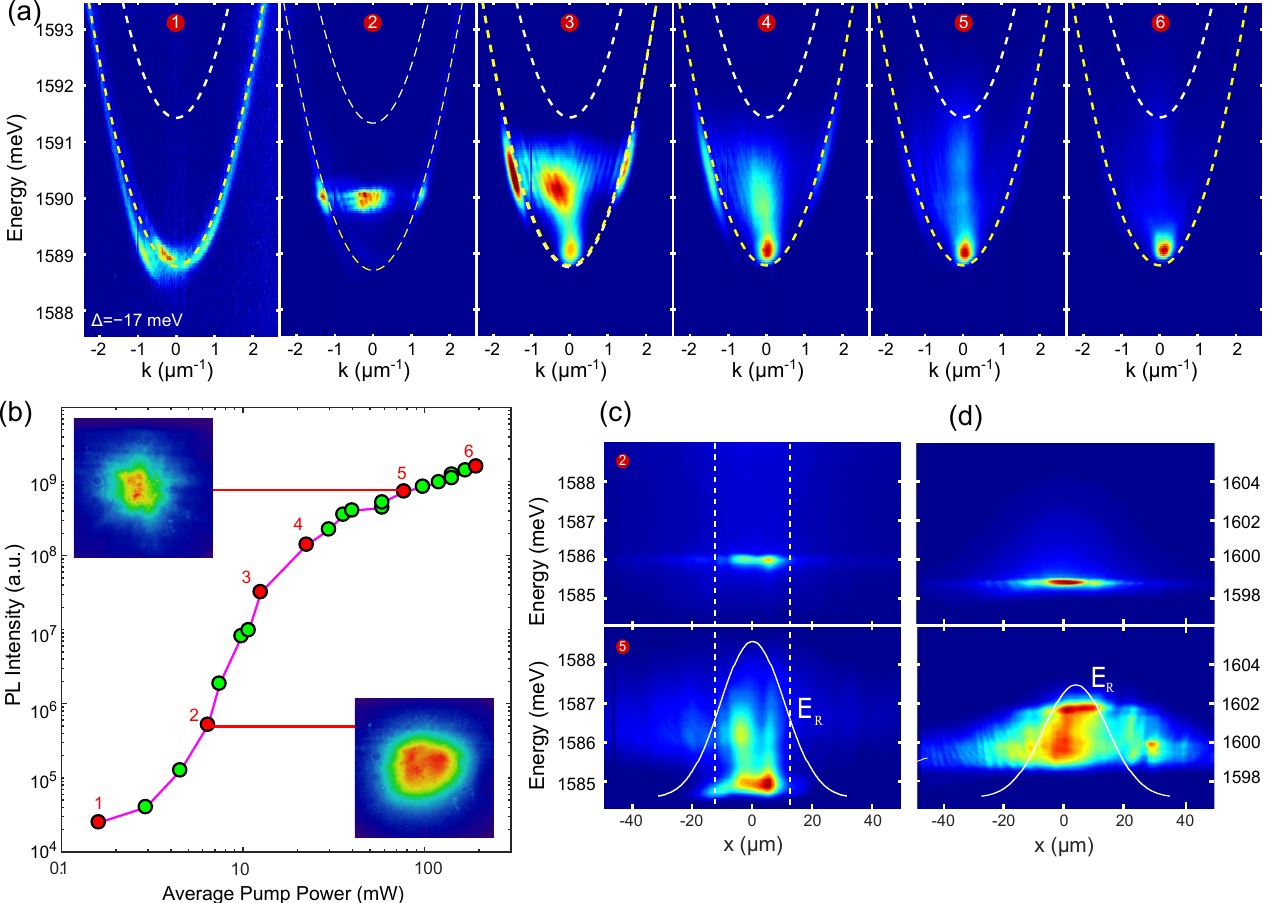}
\caption{(a) Dispersion of lower polaritons $E(k\equiv k_x)$ averaged over $10^6$ realisations of the pulsed experiment (1-2) below, (3) near, and (4-6) above threshold for the large negative exciton-photon detuning, $\Delta=-17$ meV. Top (bottom) dashed lines correspond to the cavity photon (lower polariton), respectively. (b) Measurements of the emission intensity below and above condensation threshold. Insets show time-integrated real-space images of the polariton density below and above threshold in a $70\times70$ $\mu$m detection window. (c) Lower polariton energy vs position ($x$) for low (top) and high (bottom) excitation powers corresponding to the dispersion shown in (a). Vertical dashed lines correspond to FWHM of the optical excitation pulse, and thin curves mark the blue shift due to the reservoir density, $E_R$ deduced from the theoretical model. (d) Same as in (c), but for more excitonic polaritons at $\Delta=-3$ meV. The difference in the energy scales is due to the lower powers required to achieve condensation \cite{Bloch_APL_2009} and overall blue shift of the LP dispersion.}
\label{power_series}
\end{figure}

The transition to condensation in our experiment is driven by the far-off-resonant Gaussian pump with the FWHM of $\sim 25 \mu$m. The pumping is performed by a short ($\sim 140$ fs) laser pulse (see Methods). Polariton dispersion and real-space spectra characterising the transition are shown in Fig. \ref{power_series} (a,b), where Figs. \ref{power_series}(a3-a6) demonstrate transition to condensation at $E_{\rm min}({\bf k}=0)$ with increasing rate of injection of the free carriers (optical pump power). Figure \ref{power_series} is representative of the condensation process when the detuning between the cavity photon energy and the exciton resonance, $\Delta=E_c-E_{ex}$, is large and negative, i.e. for the polaritons that have a larger photonic fraction. Results for other values of detuning are presented in Supplementary Information (SI). It should be noted that these images represent polariton dynamics averaged over $10^6$ realisations of the polariton condensation experiment in the pulsed regime. Each of the single realisation images is also time-integrated since the camera collects the cavity photoluminescence throughout the entire lifecycle ($\sim 200$ ps) of the condensate (see Methods). This integration over the duration of the single-shot experiment should be taken into account when interpreting the $E(k)$ and $E(x)$ images in Figure \ref{power_series}. The position of a maxima in $x$ or $k$ in these images will correspond to the maxima of the polariton density, and the photoluminescence collected during the process of energy relaxation and decay of the condensate will lead to `smearing out' of the image along the $E$-axis.

\begin{figure}[ht]
\centering
\includegraphics[width=10cm]{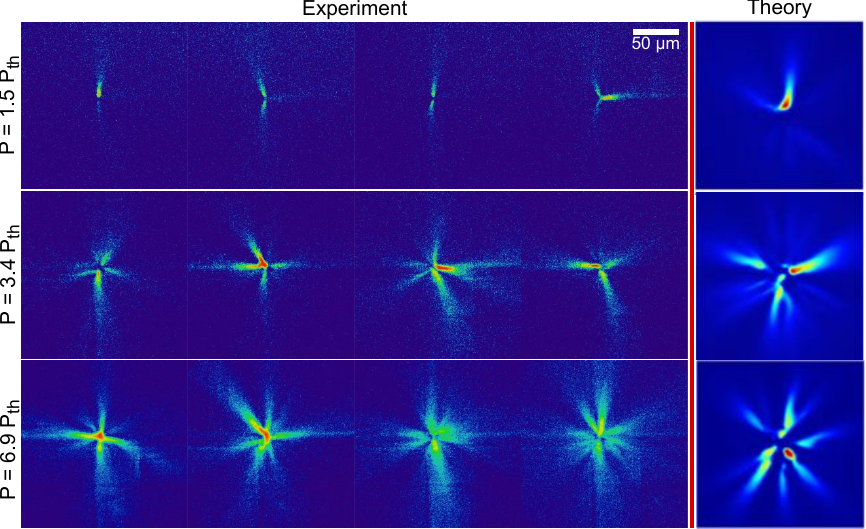}
\caption{Single-shot real space images of the photoluminescence intensity (proportional to polariton density) above condensation threshold $P_{th}\approx10$ mW, in the far red-detuned (large negative detuning, $\Delta=-22$ meV) regime for various pump powers. Each panel represents a single realisation of a spontaneous condensation process. The intensity is plotted on a log scale for the experimental images to elucidate the details of regions with low density emission. Far right column shows corresponding images obtained by numerical modelling.}
\label{sh_realisations}
\end{figure}

Near the condensation threshold, we observe formation of a high-energy state shown in Fig. \ref{power_series}(a2,a3) characterised by a low density (as confirmed from low levels of photoluminescence intensity in Fig. \ref{power_series}(b)) and a flat dispersion. At the first glance, the transition from this highly-non-equilibrium state at high energies to a ${\bf k}=0$ condensate at the bottom of the LP dispersion in Fig. \ref{power_series} can be attributed entirely to the energy relaxation processes. As has been previously demonstrated \cite{two_regimes} for the negative detuning between cavity photon and exciton resonances, i.e. for the polaritons with a high photonic fraction, energy relaxation of polaritons down the LP branch is inefficient due to reduced scattering with phonons. Under {\em cw} excitation conditions, this leads to accumulation of polariton density in non-equilibrium metastable high-$k$ states leading to stimulated bosonic scattering into these modes. In this `kinetic condensation' regime, condensation into high-energy, high in-plane momenta states ($k_{||}\neq 0$) is typically observed \cite{LeSiDand05,Bloch_APL_2009}. In contrast, in the regime of near-zero and positive detuning, $\Delta>0$, highly efficient phonon-assisted relaxation leads to efficient thermalization and high mode occupations near the minimum of the LP branch $E(k)$. Subsequently, condensation occurs into the ground state $k=0$ assisted by enhanced bosonic scattering due to strong interactions of highly excitonic polaritons \cite{Bloch_APL_2009,two_regimes}.

In our non-stationary condensation regime, the polaritons created by a short pulse just above the threshold power accumulate on top of the potential hill provided by the incoherent reservoir, which defines the offset (blue shift) of this state relative to $E_{\rm min}({\bf k}=0)$. The ${\bf k}\neq0$ tails on the polariton dispersion arise due to ballistic expansion and flow of polaritons down the potential hill \cite{Wouters2008}. In the absence of appreciable phonon-assisted energy relaxation, this flow is non-dissipative and the initial blue shift is converted into the kinetic energy. Similar behaviour has been described in previous experiments in {\em cw} regime \cite{ballistic,Sanvitto_arXiv}. These experiments also demonstrated that, as the pump power grows, phonon-assisted relaxation into the ground state increases, leading to the transition to the energy and momentum ground state. Importantly, in the {\em cw} regime the ground state condensate forms at the bottom of the potential hill formed by the pump-induced reservoir and is therefore {\em spatially offset} from the pump region  \cite{Sanvitto_arXiv}. The blue shift of this state from the minimum of the LP dispersion is purely due to polariton-polariton interaction, and is negligible for weakly interacting photon-like polaritons at large negative detunings. In striking contrast to these {\em cw} observations, our measurements of the condensate energy in real space $E(x)$ performed in the pulsed regime and shown in Fig. \ref{power_series}(c) reveal that the condensate forms in the spatial region {\em completely overlapping} with the long-lifetime reservoir. The same behaviour is observed for more excitonic polaritons at small negative detunings \ref{power_series}(d). The absence of the reservoir-induced blue shift above condensation threshold for highly photonic polaritons is therefore puzzling and can be understood only by analysing the intricate details of the single-shot condensation dynamics, as discussed below.


\begin{figure}[ht]
\centering
\includegraphics[width=10cm]{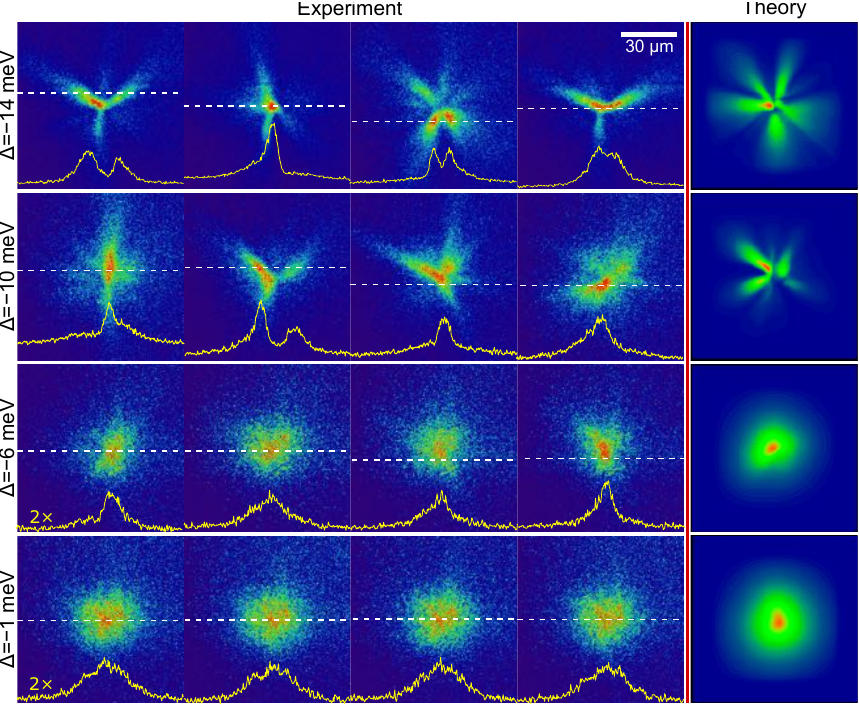}
\caption{Single-shot real space images of the photoluminescence intensity (proportional to polariton density) for $P/P_{th}\sim6$ and varying values of detuning $\Delta$. Each panel represents a single realisation of spontaneous condensation and includes an image of the intensity profile at a cross-section marked by a dashed line. Intensity is plotted on linear scale. The far right column shows corresponding images obtained by numerical modelling.}
\label{sh_detuning}
\end{figure}

\subsection{Single-shot imaging.}

A single-shot real-space imaging of the condensate dynamics captures the photoluminescence produced by polaritons in the microcavity throughout their life cycle after the excitation by a single off-resonant laser pulse. The single-shot regime is ensured by a pulse-picking technique (see Methods). For the Gaussian excitation spot, we observed strong filamentation of the condensate at highly negative detunings ($\Delta<-14.5$~meV) with filaments extending for macroscopic distances $>100$~$\mu$m, as seen in Fig. \ref{sh_realisations}. The orientation of the filaments varies from shot-to-shot (see SI for more detail), which rules out pinning of the condensate by a disorder potential in the microcavity \cite{multimode2}. Remarkably, filamentation of the condensate persists even when the statistically averaged dispersion shows {\em clear spectral signatures of the Bose-Einstein condensation in a true ground state} of the system [Fig. \ref{power_series} (a3-a6)], and the time-averaged image of the spatial density distribution displays a smooth, spatially homogeneous profile [Fig. \ref{power_series} (b), inset].

With increasing detuning, hence a larger excitonic fraction, we observe a transition from strong filamentation to a Gaussian-like condensate profile with reduced shot-to-shot density fluctuations. This transition is clearly seen in the real space images of the condensate presented in Fig. \ref{sh_detuning}. Appreciable blue shift of the ground state due to polariton-polariton interactions is also seen for detunings $\Delta>-3$ meV well above condensation threshold, $P/P_{th}>5$, however at this stage the prevalence of the strong coupling regime in the excitation region cannot be assured at the early stage of the single-shot dynamics due to the large density of the carriers injected by the initial pulse \cite{weak_coupling_1,weak_coupling_2,weak_coupling_3}. 

\begin{figure}[ht]
\centering
\includegraphics[width=16cm]{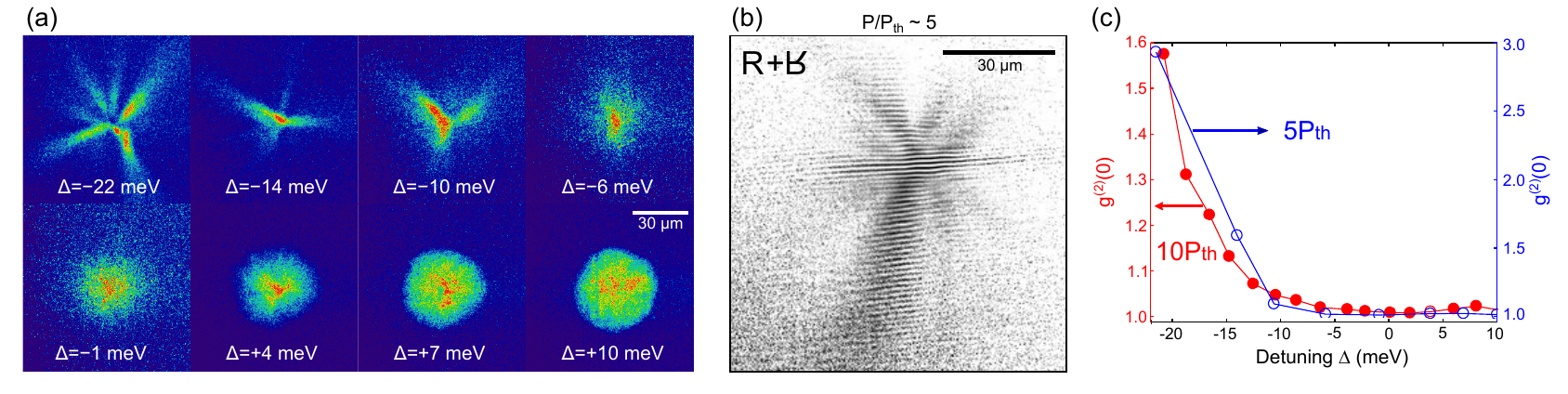}
\caption{(a) Single-shot real space images of the polariton density above the condensation threshold, $P/P_{th}\sim 5$, for a range of detuning values, $\Delta$. (b) Interference of the condensate emission with its retroreflected image demonstrating the range of first-order spatial coherence, $g^{(1)}$, extending across the entire condensate despite strong filamentation. The image is taken at $\Delta=-22$ meV and $P/P_{th}\sim 5$. (c) Measurements of the second-order spatial density correlation function $g^{(2)}(0)$ for $P/P_{th}\sim 5$ and $P/P_{th}\sim 10$. Values of $g^{(2)}(0)>2$ indicate non-Gaussian distribution of fluctuations.}
\label{sh_g2}
\end{figure}

The measure of the phase fluctuations in the system and therefore an indication of the long-range order (or absence thereof) is given by the spatial first-order correlation function $g^{(1)}$, which can be deduced from the interference of the polariton emission in a single shot. The typical single-shot interference pattern shown in Fig. \ref{sh_g2}(b) for the highly photonic condensate with high degree of filamentation, shows that the spatial coherence extends across the length of $\sim 100$ $\mu$m, which is comparable to the size of the whole condensate (details of this measurement are found in SI). Remarkably, this conclusion holds even for low pump powers above threshold, where only few filaments are formed, and the condensate is highly spatially inhomogeneous.


To quantify the transition to a condensed state with reduced shot-to-shot density fluctuations, we calculate the zero-time-delay density second-order correlation function:
\begin{equation}
 g^{(2)}(\delta x,\delta y)=\frac{\langle I(x,y)I(x+\delta x,y+\delta y)\rangle}{\langle I(x,y)\rangle\langle I(x+\delta x,y+\delta y)\rangle}, 
\end{equation}
where $I(x,y)$ is the camera counts at pixel position $(x,y)$ of a single-shot image, and $\langle\rangle$ represents the ensemble average over the number of experimental realisations. This function is a measure of density fluctuations in the condensate. Since the single-shot images are time-integrated, the experimentally measured $g^{(2)}$ function is a weighted average over the lifetime of the condensate. For second-order coherent light-matter waves $g^{(2)}(0,0)\equiv g^{(2)}(0)=1$, and for the condensate with strong density fluctuations $g^{(2)}(0)> 1$ \cite{Blakie}. 

The measurement of $g^{(2)}(0)$ in our experiment is presented in Fig. \ref{sh_g2}(c), and demonstrates a clear transition from second-order incoherent polaritons for $\Delta<-5$~meV to second-order coherent polariton BEC regime. The $g^{(2)}(0)>1$ indicates that condensates of weakly-interacting photon-like polaritons are characterised by large statistical fluctuations, which nevertheless coexist with macroscopic phase coherence as evidenced by the $g^{(1)}$ measurement. Earlier experiments with short-lifetime polaritons support this conclusion \cite{vortex_pair}. Similar behaviour was recently observed for photon condensates strongly coupled to a hot reservoir, which acts both as a source of particles and a source of thermal fluctuations \cite{weitz2016} thus realising the grand-canonical statistical conditions. The apparent drop of $g^{(2)}(0)\to 1$ for condensates of more excitonic particles at larger detuning values then primarily indicates growth of coherent condensate fraction in the system and depletion of the reservoir, as well as suppression of fluctuations due to increased interactions \cite{interactions,Wouters2009}. 


\subsection{Theoretical modeling.}

To model the formation and decay of the condensate produced by a single laser pulse, we employ the open-dissipative Gross-Pitaevskii model \cite{Wouters2007} with a phenomenological energy relaxation responsible for the effective reduction of the chemical potential of the condensate and an additional stochastic term accounting for fluctuations \cite{Wouters2008}:
\begin{eqnarray}
i\hbar\frac{\partial\psi(\mathbf{r})}{\partial t}&=&\left[ (i\beta-1)\frac{\hbar^2}{2m}\nabla^2+g_c|\psi|^2+g_Rn_R+i\frac{\hbar}{2}\left(Rn_R-\gamma_c\right)\right]\psi(\mathbf{r}) + \frac{dW}{d t}, \label{SEq1}
\\
\frac{\partial n_R(\mathbf{r})}{\partial t}&=&-(\gamma_R+R|\psi(\mathbf{r})|^2)n_R(\mathbf{r})+P(\mathbf{r}).
\label{SEq2}
\end{eqnarray}
In Eq. (\ref{SEq1}), $R$ defines the stimulated scattering rate, $\gamma_c$ and $\gamma_R$ are the decay rates of condensed polaritons and excitonic reservoir, correspondingly. Constants $g_c$ and $g_R$ characterise the strengths of polariton-polariton and polariton-reservoir interactions, respectively. The rate of injection of the reservoir particles, $P(\mathbf{r})$, in Eq. (\ref{SEq2}) is proportional to the pump power, and its spatial distribution is defined by the pump profile.

The model equations in this form can be consistently derived within the truncated Wigner approximation~\cite{Wouters2009,Davis_review}. The term $dW/dt$ introduces a stochastic noise in the form of a Gaussian random variable with the white noise correlations: 
\begin{equation}
\langle dW^{*}_idW_j\rangle= \frac{\gamma_c+R n_R(\textbf{r}_i)}{2(\delta x \delta y)^2} \delta_{i,j}dt, \qquad \langle dW_idW_j\rangle=0, 
\end{equation}
where $i$, $j$ are discretization indices: $\mathbf{r}_i=(\delta x$,$\delta y)_i$. We note that both the loss and the gain, $\gamma_c$ and $R n_R$, contribute to this term~\cite{Wouters2008,Wouters2009}. The single-shot realisation of the spontaneous condensation experiment thus corresponds to the {\em single realisation of the stochastic process} modeled by equations (\ref{SEq1},\ref{SEq2}).

Importantly, the model parameters are varied consistently with the characteristic values for long-life polaritons at various values of the exciton-photon detuning. Specifically, we can estimate the values of the interaction coefficients $g_c$ and $g_R$ from the corresponding nonlinear part of the photon-exciton interaction Hamiltonian re-written in the basis of the lower and upper polariton states, $\hat{\psi}_{LP}=C\hat{\phi}+X\hat{\chi}$ and $\hat{\psi}_{UP}=X\hat{\chi}-C\hat{\phi}$, where $C$ and $X$ are the real-valued Hopfield coefficients \cite{CiutiREV13}. When the cavity is excited by linearly polarized light (see, e.g., Ref.~\cite{bistability}), $g_{c}=g_{\rm ex} |X|^4$ and $g_{R}=g_{\rm ex} |X|^2$. Here $g_{ex}=(\alpha_1+\alpha_2)/2$ is exciton-exciton interaction strength, which is the sum of the triplet and singlet contributions (typically $\alpha_2 \ll \alpha_1$), and we have neglected the saturation of the exciton interaction strength \cite{CiutiREV13}. The absolute value of the triplet interaction coefficient is difficult to determine and is debated \cite{Nelson15}. Here we assume the standard value $\alpha_1=6E_0a^2_B$, where $E_0$ is the binding energy of the Wannier-Mott exciton, and $a_B$ is the exciton Bohr radius in the particular semiconductor \cite{ciuti98,tassone}. For GaAs QW microcavities used in our experiments $a_{B}\approx 7$ nm and $E_0\sim 10$ meV. The Hopfield coefficient, which defines the value of the excitonic fraction, depends on the exciton-photon detuning as follows: $|X|^2=(1/2)\left[1+\Delta/\sqrt{4\hbar^2\Omega^2+\Delta^2}\right]$, where $2\hbar\Omega$ is the Rabi splitting. Furthermore, the LP effective mass and decay rate for polaritons are also detuning-dependent via the Hopfield coefficients: $1/m=|X|^2/m_{\rm ex}+(1-|X|^2)/m_{\rm ph}$, $\gamma_c=|X|^2\gamma_{\rm ex}+(1-|X|^2)\gamma_{\rm ph}$, which affects the respective parameters in the model equation. Finally, we assume that the stimulated scattering rate from the reservoir into the polariton states is more efficient for more excitonic polaritons: $R\sim R_0|X|^2$. 

The phenomenological relaxation coefficient, $\beta$, defines the rate of the kinetic energy relaxation due to the non-radiative processes, such as polariton-phonon scattering, and is critical for modeling the highly non-equilibrium, non-stationary condensation dynamics presented here. This parameter is usually assumed to depend on the polariton \cite{Wouters_relaxation} or reservoir \cite{Liew_ring} density, however we find that the effect of the increasing detuning (from negative to positive) on growing efficiency of energy relaxation towards low-momenta states in our experiment is adequately described by increasing the value of the relaxation constant $\beta$.

The excellent agreement between the numerical simulations and experiment can be seen in real-space images shown in Figs. \ref{sh_realisations} and \ref{sh_detuning}. Importantly, the filamentation efffect observed in the experiments is reproduced in numerical simulations using the detuning-dependent parameters as described above. It is also critical to note that the initial condition for the simulations is the white noise $\psi_0$, which essentially ensures that a non-stationary polariton mean-field inherits strong density and phase fluctuations \cite{Liew15}, because neither the reservoir nor the polariton density reach a steady state in our experiments. 

\begin{figure}[ht]
\centering
\includegraphics[width=16cm]{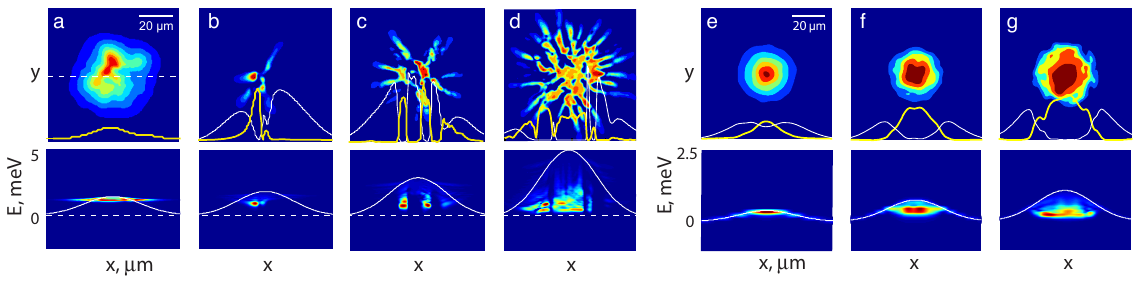}
\caption{Numerically calculated (top) single-shot real space images and (bottom) real space spectra $E(x)$ at and above the condensation threshold, for (a-d) highly photonic polaritons ($\Delta=-20$ meV) and (e-f) more excitonic polaritons ($\Delta=-3$ meV). White and yellow lines in the top row are the $y=0$ cross-sections of the reservoir ($n_R$) and condensate ($|\psi|^2$) densities, respectively. Panel (a) shows the condensate density at threshold and is scaled up by a factor of $30$; the reservoir density is off the scale and is not shown. In the bottom row, the white solid curve corresponds to the initial blue shift, $E_R$, due to non-depleted reservoir. Zero on the energy axis corresponds to the minimum of the LP dispersion for the particular value of the detuning. The ratio $P/P_{th}$ is equal to (b) 3, (c) 6, (d) 10, and (e) 1.2, (f) 5, (g) 8. The spectra shown in the bottom panels represent time-integrated images over the life cycle of the condensate in a single-shot experiment without averaging over multiple condensate realisations (in contrast to Fig. \ref{power_series}(c)).}
\label{theory}
\end{figure}

The model (\ref{SEq1},\ref{SEq2}) allows us to simulate the process of the condensate formation and dynamics after an initial density of reservoir particles is injected by the pump. Importantly, using this model we can not only reproduce the spatial and spectral signatures of the condensate in the different regimes shown in Figs. \ref{sh_realisations}, \ref{sh_detuning}, but also to trace the process of the spontaneous condensation near threshold, where the densities are too low to be captured by single-shot imaging in our experiments. The simulations show that the condensate formation is seeded in one or several randomly located hot spots, which then locally deplete the reservoir at the spatially inhomogeneous rate proportional to the condensate density and the stimulated scattering rate $\gamma_{RD}=R|\psi|^2$. This depletion remains insignificant just below and at threshold, which leads to the bulk of polariton emission originating from the high-energy states on top of the reservoir-induced potential hill, which are blue-shifted from $E_{\rm min}({\bf k}=0)$ by the value of $E_R=g_R n_R$ [see Fig. \ref{theory} (a)] and result in the emission shown in Figs. \ref{power_series}(a2). Once the condensate forms at the particular hot spot(s), the $\gamma_{RD}$ at this location dramatically increases, and the condensate becomes trapped in local reservoir-induced potential minima, leading to spatial filamentation seen in Fig. \ref{sh_realisations} and Fig. \ref{theory}. The location and size of the local trapping potentials is random at each realisation of the condensation process, which leads to large shot-to-shot density variations. We stress that this dramatic `hole-burning' effect due to irreversible reservoir depletion is not possible in a {\em cw} excitation experiment where the reservoir is continuously replenished by a pump laser. Instead, condensation at ${\bf k}=0$ occurs in a trapped dissipative state formed in a gain-guiding potential \cite{gain_guiding_exp,gain_guiding_th}, and is massively blue-shifted by the reservoir \cite{Wouters2008}.

As the excitonic fraction in a polariton increases with growing detuning, phonon-assisted energy relaxation becomes more efficient and tends to suppress high-momentum excitations. This tendency is well captured by the phenomenological relaxation in our model, which provides damping of the (spatial) spectral components at the rate $\gamma_{\rm ER}\sim \hbar\beta |k|^2/m$. The suppression of high-$k$ fluctuations of density results in larger hot spots and a larger area of reservoir depletion right at the onset of condensation. This leads to the formation of condensates without filamentation and, at high pump powers, complete phase separation between the condensate and the reservoir due to the depletion process. The lack of spatial overlap between the condensate and the thermal reservoir leads to reduced statistical fluctuations \cite{Wouters2009}, as observed in Fig. \ref{sh_g2}(c). We also note that time-resolved measurements of $g^{(2)}(0)$ during the condensate formation in a CdTe microcavity show transition from a thermal to coherent emission during single-shot dynamics \cite{deveaud_g2}. Since our measurement is integrated over the duration of the condensate life cycle, a value close to $g^{(2)}(0)=1$ indicates that the condensate quickly reaches a coherent stage and remains coherent as it decays. Due to the largely depleted reservoir, no revival of the thermal emission occurs as the condensate decays.

\section*{Discussion} 

The remarkable agreement between our theory and single-shot experimental results unambiguously links transition to the spontaneous condensation in low energy and momenta states to the combination of two processes: {\em energy relaxation}, represented in our model by the rate $\gamma_{\rm ER}$ and local {\em reservoir depletion} characterised by $\gamma_{\rm RD}$. As long as both of these rates are greater than the rate of the polariton decay $\gamma_c$, the hole-burning effect and efficient energy relaxation drive the condensation to the ground state. Near threshold, the condensate growth rate $\gamma_{cg} \sim Rn_R^0$, where $n_R^0$ is the initial density of the reservoir injected by the excitation pulse, is also competing with the rate of ballistic expansion of polaritons due to the interaction with reservoir. The latter is determined by the velocity of the polaritons acquired as the interaction energy with the reservoir $E_R=g_R n_R$ is converted to kinetic energy, and can be estimated as $\gamma_{exp}\sim (1/L)\sqrt{2E_R/m}$, where $L$ is the spatial extent of the pump-reservoir region. Inefficient energy relaxation, and fast growth of the condensed fraction, $\gamma_{cg}>\gamma_{\rm exp}$, leads to large density fluctuations and condensation in several spatially separated filaments driven by the condensate depletion. The above scenario is realised in our experiment for large negative detuning, i.e., for largely photonic polaritons. In contrast, efficient energy relaxation and lower reservoir densities (pump powers) required for condensation lead to the rates of ballistic expansion being comparable to that of condensate growth, which results in more homogeneous condensate density. This scenario is realised for more excitonic polaritons at small negative and near-zero detunings.


Our results have several important implications for polariton physics. First, they offer a striking demonstration of the strong role of the reservoir depletion on the formation of a polariton condensate in the non-equilibrium, non-stationary regime. This demonstration is uniquely enabled by the single shot nature of our experiment which ensures that, once depleted, the reservoir is never replenished: the reservoir which feeds the polariton condensate is created by the laser pulse and is depleted and/or decays before the next excitation pulse arrives. For polaritons with a high admixture of particle (exciton) component, i.e. in the regime when both reservoir depletion and phonon-assisted energy relaxation dominate the condensation dynamics, we are therefore able to create high-density condensates that are {\em spatially separated from the reservoir}, the latter acting as both a source of polaritons and a source of strong number (density) fluctuations. Such condensates, apart from demonstrating macroscopic phase coherence, exhibit second-order spatial coherence.

Secondly, our results indicate that spatial filamentation, similar to that attributed to dynamical instability of the polariton condensate in an organic microcavity \cite{Bobrovska2016}, is an inherent feature of the non-stationary polariton condensation process, which is completely masked by any statistically-averaging measurement and can only be uncovered in the single-shot regime. The filamentation arises due to the random spatial fluctuations inherited from the incoherent reservoir at the onset of the condensation. The efficient energy relaxation is critical for gradual suppression of these fluctuations with growing exciton-photon detuning, and subsequent formation of condensates with reduced degree of filamentation. We note that our theoretical model reproduces the typical filamentation behaviour observed in experiments {\em without the assumption that the model parameters lie within the dynamical (modulational) instability domain}, i.e. when the condensate would be stable to modulations of density. It is still possible that modulational instability does affect the dynamics of the condensate, however in our experiments we have not observed any sharp threshold behaviour typically associated with an onset of instability \cite{Hulet,Robins,Turitsyna2}.


Last but not least, the single-shot imaging technique is a powerful tool for further studies of fundamental properties of the non-equilibrium condensation process, such as development of the macroscopic phase coherence in a polariton condensate strongly coupled to the reservoir. In particular, combination of the first-order correlation measurements and direct imaging of phase defects in the single-shot regime could assist in testing the Kibble-Zurek-type scaling laws in open-dissipative quantum systems \cite{Liew15}. 


\begin{methods}

\subsection{Sample.} The high Q-factor microcavity sample used in this work consists of four $7$-nm GaAs quantum wells embedded in a $3\lambda/2$ microcavity with distributed Bragg reflectors composed of $32$ and $40$ pairs of Al$_{0.2}$Ga$_{0.8}$As/AlAs $\lambda/4$ layers; similar to the one used in Ref.~\cite{ballistic}. The Rabi splitting is $\Omega = 14.5$ meV, and the exciton energy at normal incidence pumping is $E_X({\bf k}=0) = 1608.8$ meV. The effective mass of the cavity photon is $m_{ph}\approx 3.9\times10^{-5} m_e$, where $m_e$ is a free electron mass.

\subsection{Experiment.} We used a photoluminescence microscopy setup typical of off-resonant excitation experiments with exciton-polaritons. A single $50\times$ objective (NA$=0.5$) is used to focus the pump laser and collect the photoluminescence from the sample. The single-shot imaging is realised by employing a high contrast ratio ($\sim1:10,000$) pulse picker that picks a single $140$ fs pulse at $<100$ Hz rate from a $80$ MHz mode-locked Ti:Sapphire laser (Chameleon Ultra II). It is synced to an Electron-Multiplying CCD camera (Andor iXon Ultra 888) which is exposed for at least $10$ $\mu$s before and after the pulse. The camera therefore records photoluminescence from the sample, which is integrated over the entire lifetime of the condensate and reservoir during a single realisation of the condensation experiment. The excitation energy is tuned far above ($\sim 100$ meV) the polariton resonance to ensure spontaneous formation of the polariton condensate. Each acquired image is a time-integrated (over the whole emission lifetime) real-space distribution of polaritons in a single realisation of the condensation experiment. All other experimental images were taken without pulse picking, hence averaging over $10^6$ pulses. 



\end{methods}



\begin{addendum}
\item This work was supported by the Australian Research Council. D.W.S. acknowledges support from the U.S. Army Research Office project W911NF-15-1-0466.
\end{addendum}


\end{document}